\documentstyle[twoside,fleqn,espcrc2,epsf]{article}

\def\gtwid{\raise.3ex\hbox{$>$\kern-.75em\lower1ex\hbox{$\sim$}}}
\def\ltwid{\raise.3ex\hbox{$<$\kern-.75em\lower1ex\hbox{$\sim$}}}

\def\ie{{\it i.e.},\ }
\def\eg{{\it e.g.},\ }
\def\et{{\it et al.}}

\def\cO{{\cal O}}

\def\nsection#1 #2{\leftline{\rlap{#1}\indent\relax #2}}

\def\prl#1{Phys.\ Rev.\ Lett.\ {\bf #1}}
\def\prd#1{Phys.\ Rev.\ {\bf D#1}}
\def\plb#1{Phys.\ Lett.\ {\bf #1B}}

\def\boulder{Nucl.\ Phys.\ {\bf B} (Proc.\ Suppl.) {\bf 73} (1999)}

\newcommand{\AmS}{{\protect\the\textfont2
  A\kern-.1667em\lower.5ex\hbox{M}\kern-.125emS}}

\newcommand{\fBs}{$f_{B_s}$}

\title{Heavy-Light Decay Constants 
with Dynamical Gauge Configurations 
and Wilson or Improved 
Valence Quark Actions}

\author{ C.~Bernard,\hskip-0.03in
\address{{\vskip-0.10in{\hskip 0.07in Department of Physics, Washington
University, St.~Louis, MO 63130, USA}}} 
T.~DeGrand,\hskip-0.03in
\address{Physics Department, University of Colorado, Boulder, CO 80309, USA} %
C.~DeTar,\hskip-0.03in
\address{Physics Department, University of Utah, Salt Lake City, UT 84112, USA}
Steven~Gottlieb,\hskip-0.03in
\address{Department of Physics, Indiana University, Bloomington, IN 47405, USA}
\hskip-0.03in\thanks{presented by S.\ Gottlieb}
Urs~M.~Heller,\hskip-0.03in
\address{SCRI, Florida State University, Tallahassee, FL 32306-4130, USA} 
J.~Hetrick,\hskip-0.03in
\address{Department of
Physics, University of the Pacific, Stockton, CA 95211, USA} 
C.~McNeile,\hskip-0.03in
\address{Dept.\ of Math Sci., University of Liverpool, Liverpool, L69 3BX, UK}
K.~Orginos,\hskip-0.03in
\address{Department of Physics, University of Arizona, Tucson, AZ 85721, USA} 
R.~Sugar
\address{Department of Physics, University of California, Santa Barbara, CA
93106, USA} 
and
D.~Toussaint$\!\null^{\rm h}$
\advance\baselineskip -2pt
} 

\begin{document}

\begin{abstract}
We describe a calculation of heavy-light decay constants including
virtual quark loop effects.
We have generated 
dynamical gauge configurations 
at three $\beta$ values 
using  two flavors of Kogut-Susskind quarks with a range of masses.
These are analyzed with a Wilson valence quark action. 
Preliminary results based on a ``fat-link'' clover valence quark action
are also reported.
Results from the two methods differ by $30$ to $50$ MeV, which
is presumably due to significant --- but as yet unobserved ---
lattice spacing dependence in one or both of the approaches.

\end{abstract}

\maketitle

Decay constants for the $B$ and $B_s$ mesons are crucial for the accurate
determination of the CKM mixing matrix.
Reference~\cite{MILC-PRL} describes our evaluation 
of  these decay constants in the quenched approximation;
the results are consistent with those from several other groups \cite{OTHERS}.
The effects of quenching in \cite{MILC-PRL} were estimated by comparing 
with results including dynamical quark
effects at fixed lattice spacing. 
We now have enough results with $N_F=2$ dynamical quarks
to start to study the continuum limit in
the dynamical theory. This is the crucial step to go from
quenched answers with estimates of quenching effects to
true dynamical answers.

Dynamical gauge configurations have been generated with two flavors
of staggered quarks at $\beta=5.445$, $5.5$, and $5.6$, with a range
of dynamical masses.  (See Table \ref{tab:lattices}.)
We have analyzed each set with Wilson valence quarks (both heavy and light)
as well as static heavy quarks, as in \cite{MILC-PRL}.  In addition,
we have begun to use
heavy and light ``fat-link'' \cite{FATREFa,FATREFb} 
clover valence quarks.  

\begin{table}
\caption{Lattice parameters.  All sets use $N_F=2$ dynamical
staggered quarks and are analyzed with Wilson valence
quarks. To date, 98 configurations of set R 
have been analyzed with fat-link clover valence quarks.
Set G was generated by HEMCGC.}
\label{tab:lattices}
\centering
\begin{tabular}{ccccc} \hline
name& $\beta$ & $am_q$ &size &\# configs. \\
\vrule height 10pt width 0pt L& $ 5.445$ &$0.025 $&  $ 16^3 \times 48$& 100\cr
\hline
\vrule height 10pt width 0pt N& $ 5.5$ &$0.1 $&  $ 24^3 \times 64$&    100 \cr
\hline
\vrule height 10pt width 0pt O& $ 5.5$ &$0.05 $&  $ 24^3 \times 64$&    100 \cr
\hline
\vrule height 10pt width 0pt M& $ 5.5$ &$0.025 $&  $ 20^3 \times 64$&   199 \cr
\hline
\vrule height 10pt width 0pt P& $ 5.5$ &$0.0125 $&  $ 20^3 \times 64$&   199 \cr
\hline
\vrule height 10pt width 0pt U& $ 5.6$ &$0.08 $&   $24^3 \times 64$&    201 \cr
\hline
\vrule height 10pt width 0pt T& $ 5.6$ &$0.04 $&   $24^3 \times 64$&    202 \cr
\hline
\vrule height 10pt width 0pt S& $ 5.6$ &$0.02 $&   $24^3 \times 64$&    201 \cr
\hline
\vrule height 10pt width 0pt G& $ 5.6$ &$0.01 $&   $16^3 \times 32$&    200 \cr
\hline
\vrule height 10pt width 0pt R& $ 5.6$ &$0.01 $&   $24^3 \times 64$&    200 \cr
\hline
\hline
\vspace{-30pt}
\end{tabular}
\end{table}

In the fat-link clover case, 
we implement the full Fermilab program \cite{EKM} through
$\cO(a)$ and through $\cO(1/M)$, 
including the 3-dimensional rotations (``$d_1$'' terms).
The shift to the kinetic mass is done as in the 
Wilson case \cite{MILC-PRL}, except
that tadpole improvement is not needed.

In general, we treat the dynamical quark configurations as
fixed backgrounds and perform chiral extrapolations in the valence
quark mass only; \ie we do ``partial quenching.'' However, we
also try extrapolating with $m_{\rm valence} = m_{\rm dynamical}$ 
(``full unquenching'').
The difference 
is
treated as a systematic error, although in most cases it is
smaller than the statistical errors.

Most other systematic errors (excited states, chiral extrapolation, 
fitting errors in 1/M, perturbation theory [in Wilson case], 
difference between $m_2$ amd $m_3$ [in Wilson case])
are estimated the same way as in the quenched approximation \cite{MILC-PRL}.

Finite volume errors are estimated by comparing results
of sets G and R. Since set G has a smaller physical volume than
all other runs, this is an overestimate.

The hardest errors to control with our data are
discretization errors and the 
effects of omitting the dynamical strange quark.
We discuss them below.

Details about the fat-link clover approach can be found in
Refs.~\cite{LAT99_FORMFACTOR,CBTD_FAT}.
Throughout the current work, we use  $N=10$ smearing steps and 
smearing parameter $c=0.45$ ($c/6$ is the coefficient of the staple sum). 
This amount of fattening completely suppresses exceptional
configurations in the range of masses we are studying
\cite{FATREFa}. With
the standard (``thin link'') nonperturbative clover action, we found
exceptional configurations to be a very serious problem on our
perforce somewhat coarse dynamical lattices. 

The clover coefficient has been chosen equal to the tree-level
value, $C_{SW}=1$.  The fact that fattening suppresses perturbative 
corrections \cite{CBTD_FAT} 
leads us to expect that this value
should be  very close to the all-orders (in $g$) value for our
fat links.  We plan a nonperturbative
evaluation to check this.


Bernard and DeGrand \cite{CBTD_FAT} have computed
fat-link clover $Z$ factors in perturbation theory.
For
light-light (ll) and static-light (sl) $Z_A$, they find:
\vspace{-4pt}
\begin{eqnarray}
Z_A^{\rm ll} &=& 1 + {g^2 C_F\over 16 \pi^2_{\phantom{p}}}(-0.241)\cr
Z_A^{\rm sl} &=& 1 + {g^2 C_F\over 16 \pi^2}
\left(3\log\left(aM_B\right) +0.393\right)\ .
\vspace{-4pt}
\end{eqnarray}
For $Z_A^{\rm ll}$, 
$q^*=0.71/a$.
For $Z_A^{\rm sl}$, $q^*$ has not yet 
been calculated; we use the light-light $q^*$.
The mass-dependent heavy-light $Z_A^{\rm hl}$ has also not yet been computed. 
We expect that for moderately large masses, the difference between 
$Z_A^{\rm hl}$ and
$Z_A^{\rm sl}$ will be small: such finite numbers are
strongly suppressed by fattening.
We currently use $Z_A^{\rm sl}$ for heavy-lights.

At present, we have analyzed only a subset of one lattice set (R) with
fat-link clover valence quarks. 
With only two light quark masses currently available, we choose to
focus here on \fBs.

\begin{figure}[t] 
\vspace{-44pt}
\epsfxsize=1.0 \hsize
\epsffile{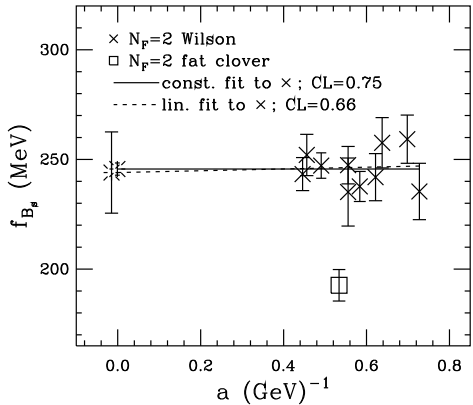}
\vspace{-28pt}
\caption{ \fBs\ {\it vs.}\ lattice spacing.
}
\vspace{-30pt}
\label{fig:fbs_dyn}
\end{figure}

Figure \ref{fig:fbs_dyn} shows \fBs\ as a function of $a$ 
in both the
Wilson and fat-link clover 
cases.
The Wilson valence points are consistent with constant behavior
in $a$;
allowing a linear term in the fit makes almost
no difference in the extrapolated value at $a=0$.  However the 
extrapolated values are 
inconsistent
with the fat-link clover result.  
Possible explanations
for this discrepancy are:


(1) The apparently constant behavior of the Wilson results is
misleading.  Indeed one expects the Wilson results on dynamical
configurations to decrease as $a\to0$ with roughly the same slope as in
the quenched Wilson case. (The quenching effect on this slope should be
roughly like the quenching effect on physical quantities,
\ie $\sim\!5$--$30\%$.) In this scenario, the reason that the
Wilson results look constant 
is that the
effects of dynamical quarks (which should raise decay constants
by deepening the potential well at $r=0$) are turning on as the lattice
spacing becomes fine enough to see the small $r$ behavior. For smaller
$a$, they would begin to fall. If we
assume that the $a=0$ limit of the Wilson data is equal to the
fat-link clover result, and that the linear slope is the
same as in the quenched case, we can make a quadratic fit to the
Wilson data with a confidence level of $0.23$ and a reasonable
quadratic term of scale $(390\, {\rm MeV})^2$.  This does not validate
the scenario, of course, but only shows that it is a consistent
possibility.

(2) Too much fattening has done violence to the physics
governing \fBs. This could be the case if, {\it e.g.}, the smearing
softens the Coulomb potential at the origin enough to reduce significantly 
the decay constants.  This would not mean that fattening is ``wrong,''
but 
that this much fattening introduces significant
lattice spacing dependence. This dependence presumably would occur
at $\cO(a^2)$ or higher, since we have argued that 
$C_{SW}=1$ is close to the nonperturbative value needed
for $\cO(a)$ improvement.

(3) Perturbation theory, used to find the 
renormalization constants in
the  fat-link case, has broken down.  This may be the case
because the fattening has so reduced the large $q$ behavior of the 
integrands that the integrals are IR dominated, and the resulting
effective coupling constant is too large. 
The small values obtained for $q^*$ are
indicative of this potential problem.
In retrospect, less fattening would have been preferable \cite{CBTD_FAT}.


The full
explanation is probably some combination of these
three scenarios.  Scenario
(1) makes it clear that, while extrapolating
the Wilson valence results with a constant may produce significant
systematic errors, it should give an {\em upper bound} to the correct result.
If scenario
(2) were the only problem with the fat-clover
data, then the fat-clover result would be a {\em lower bound} to the
correct result.  Scenario
(3) complicates the situation.
However 
it is unlikely that the correct result is much below
the fat-clover result, because it is unlikely that the Wilson 
data would have a slope much steeper than in the quenched case.  
Thus we average the constant-extrapolated Wilson 
and the fat-clover results, and use the spread 
to estimate
the discretization error.  Clearly, this analysis is 
preliminary;
much more study is needed.

With $N_F=2$, we are missing the effect of a dynamical strange quark.
To estimate
this  effect, we assume that each dynamical quark, independent
of its mass, has the same
effect on the decay constants.  This assumption is supported by the
Wilson valence data. 
The values of \fBs in Fig.~\ref{fig:fbs_dyn}, \eg
do not depend strongly on the dynamical
quark mass (which varies from $\sim\!m_s/2$ to $\sim\!4m_s$).
We thus estimate the effect of the missing strange quark by taking 
$50\%$ of the difference between the $N_F=2$ results and our older
quenched results \cite{MILC-PRL}.  

With the above caveats, our {\em preliminary} results are (in MeV
for the decay constants):
\begin{eqnarray*}
f_B \!=\! 194 (3) (22)(  {}^{+20}_{ -0});\
f_{B_s}\!\!\!\! \!&=&\!\!\!\!\! 219 (3) ({}^{+32}_{ -33})(  {}^{+25}_{ -0})
\\
f_D \!=\! 211(2) (27)({}^{+10}_{ -0});\ \
\!f_{D_s} \!\!\!\!\!&=&\!\!\!\!\! 235 (2) ({}^{+36}_{ -37} )({}^{+13}_{ -0})
\vspace{-20pt}
\end{eqnarray*}


\vspace{-20pt}
\begin{eqnarray*}
{f_{B_s}\over f_B }\!\!=\!\! 1.12 (1) (5)  ({}^{+1}_{ -2}); \quad
{f_{D_s}\over f_D}\!\! =\!\! 1.11 (0) ({}^{+4}_{ -5})  ({}^{+1}_{ -3})\ .
\end{eqnarray*}
\vspace{-6pt}

The errors are statistical,
systematic (within the $N_F=2$ ``world''),
and systematic (due to the missing dynamical strange quark), respectively.

This work was supported by the DOE and NSF.

\end{document}